\documentclass[a4paper, 12pt]{article}

\usepackage{graphicx} 
\usepackage{amsfonts} 
\usepackage{amsthm}
\usepackage{amsmath}
\usepackage{amssymb}
\usepackage{natbib}
\usepackage{lscape}
\usepackage{color}
\usepackage{subcaption}
\usepackage[left=33mm,right=27mm,top=1in,bottom=1in]{geometry} 
\usepackage{multirow}
\usepackage{authblk}

\DeclareMathOperator{\y}{\mathbf{y}}

\DeclareMathOperator{\bmu}{\boldsymbol{\mu}}

\DeclareMathOperator{\bbf}{\boldsymbol{f}}
\DeclareMathOperator{\vb}{\boldsymbol{v}}
\DeclareMathOperator{\f}{\mathbf{f}}

\DeclareMathOperator{\bz}{\mathbf{z}}

\DeclareMathOperator{\x}{\mathbf{x}}

\DeclareMathOperator{\0}{\boldsymbol{0}}

\DeclareMathOperator{\diag}{\mathrm{diag}}

\newtheorem{theo}{Theorem}
\newtheorem{lem}{Lemma}

\begin{document}
\thispagestyle{empty}

\title{Extending Owen's integral table and a new multivariate Bernoulli distribution} 

\author[1]{Marcelo Hartmann, Phd student}
\affil[1]{Department of Mathematics and Statistics, P.O. Box 68, 00014 University of Helsinki, Finland}

\date{\today} 

\maketitle

\paragraph*{Corresponding author:} 
marcelo.hartmann@helsinki.fi

\section{Description}

The table of \cite{owen:1980} presents a great variety of integrals involving the Gaussian density function and the Gaussian cumulative distribution function. For some of them analytical solution is presented and for some others, the solution is written in terms of the Owen's $T$-function \citep{owen:1980}. With little more algebra we can extend many of those equalities in his table \citep{owen:1980} and moreover a new multivariate Bernoulli distribution could be found. This extension can be useful in many practical application where quadrature methods have been applied to solve integrals involving Gaussian function and the Owen's $T$-function, e.g., \cite{jaako:2013, jarvenpaa:2017, stat:2017}.


\section{Gaussian integral I} \label{app:GaussInt}

\begin{lem} \label{lemma:2} Let $\Phi(\cdot)$ be the standard-Gaussian cumulative distribution function and $\mathcal{N}(\cdot|\mu, \sigma^2)$ the Gaussian density function with parameters $(\mu, \sigma^2) \in \mathbb{R} \times \mathbb{R}_+$. Then the following holds true,
\begin{equation} \label{GaussInt}
\int\limits_{\mathbb{R}}\prod\limits_{r = 1}^{N}\Phi\left(\dfrac{x-m_r}{v_r}\right)\mathcal{N}(x|\mu, \sigma^2)\mathrm{d}x = F_N\left(\bmu_N \vert \mathbf{m}_N, V_N \right) 
\end{equation}
where $F_N(\cdot|\mathbf{c}, \mathcal{C})$ is the $N$-dimensional Gaussian cumulative distribution function with parameters $(\mathbf{c}, \mathcal{C}) \in \mathbb{R}^N \times \mathcal{R}$, with $\mathcal{R}$ the space of positive-definite matrices (covariance matrices). Furthermore, $\mathbf{m}_N = [m_1 \cdots m_N]^T \in \mathbb{R}^N$, $\bmu_N = \mu \mathbf{1}_N \in \mathbb{R}^N$, $v_r > 0 \ \forall r$ and $V_N$ is a covariance matrix given by,
\begin{equation}
V_N = \begin{bmatrix}
v_1^2 + \sigma^2 & \ldots & \sigma^2 \\ 
\vdots & \ddots & \vdots \\ 
\sigma^2 & \ldots & v_N^2 + \sigma^2
\end{bmatrix}
\end{equation}
\end{lem}
\begin{proof}
To show \eqref{GaussInt}, start writing the left-hand side of the equation in full. Rewrite the integrand as the product of non-standard Gaussian density functions as well as the regions of integration, i.e., 
\begin{equation} \label{GaussInt1}
\int\limits_{\mathbb{R}} \int\limits_{-\infty}^x \cdots \int\limits_{-\infty}^x \prod\limits_{r = 1}^{N} \mathcal{N}(y_r|m_r, v_r^2) \mathcal{N}(x|\mu, \sigma^2)\mathrm{d}y_1 \cdots \mathrm{d}y_N \mathrm{d}x.
\end{equation}
Rewrite again using the following transformation $[x, y_1, \cdots, y_N]^T =  [w + \mu, z_1 + w + m_1, \cdots, z_N + w + m_N]^T$ and note that $|\partial (x, y_1, \cdots, y_N)/\partial (w, z_1, \cdots, z_N)| = 1$. After changing variables, group the different terms in the exponentials together to have
\begin{equation} \label{GaussInt2}
\int\limits_{\mathbb{R}} \int\limits_{-\infty}^{\mu - m_N} \cdots \int\limits_{-\infty}^{\mu - m_1} \tfrac{1}{c} \exp \left\lbrace -\dfrac{1}{2} \left[ \sum_{r = 1}^N \tfrac{(z_r + w)^2}{v_r^2} + \tfrac{w^2}{\sigma^2} \right] \right\rbrace \mathrm{d}z_1 \cdots \mathrm{d}z_N \mathrm{d}w
\end{equation}
where $c = \sigma (2\pi)^{(N+1)/2} \prod_{r=1}^N v_r$. Now, the expression inside the squared bracket is a quadratic form which is written with the following matrix form,
\begin{align} \label{quadForm}
\sum_{r = 1}^N \tfrac{(z_r + w)^2}{v_r^2} + \tfrac{w^2}{\sigma^2} =& \ w^2 \left(\sum_{r = 1}^N \tfrac{1}{v^2_r} + \tfrac{1}{\sigma^2} \right) + w \sum_{r = 1}^N \tfrac{z_r}{v^2_r} + \sum_{r = 1}^N z_r \left( \tfrac{w}{v^2_r} + \tfrac{z_r}{v^2_r} \right) \nonumber \\[0.3cm]
= & \begin{bmatrix}
w \left(\sum_{r = 1}^N \tfrac{1}{v^2_r} + \tfrac{1}{\sigma^2} \right) + \sum_{r = 1}^N \tfrac{z_r}{v^2_r} \\[0.3cm]
\tfrac{w}{v^2_1} + \tfrac{z_1}{v^2_1} \\
\vdots \\
\tfrac{w}{v^2_N} + \tfrac{z_N}{v^2_N}
\end{bmatrix}^T
\begin{bmatrix}
w \\ 
z_1 \\ 
\vdots \\
z_N
\end{bmatrix} \nonumber \\[0.3cm]
= & \begin{bmatrix}
w \\ 
z_1 \\ 
\vdots \\ 
z_N
\end{bmatrix}^T  
\begin{bmatrix}
\sum_{r = 1}^N \tfrac{1}{v_r^2} + \tfrac{1}{\sigma^2} & \tfrac{1}{v_1^2} & \cdots & \tfrac{1}{v_N^2} \\ 
\tfrac{1}{v_1^2} & \tfrac{1}{v_1^2} & \cdots & 0 \\ 
\vdots & \vdots & \ddots & \vdots \\ 
\tfrac{1}{v_N^2} & 0 & \cdots & \tfrac{1}{v_N^2}
\end{bmatrix} 
\begin{bmatrix}
w \\ 
z_1 \\ 
\vdots \\ 
z_N
\end{bmatrix} 
\end{align}
therefore \eqref{GaussInt2} is the same as
\begin{equation} \label{GaussInt3}
\int\limits_{\mathbb{R}} \int\limits_{-\infty}^{\mu - m_N} \cdots \int\limits_{-\infty}^{\mu - m_1} \tfrac{1}{c} \exp \left\lbrace  -\dfrac{1}{2} \begin{bmatrix}
w \\ 
z_1 \\ 
\vdots \\ 
z_N
\end{bmatrix}^T  
\begin{bmatrix}
\sum_{r = 1}^N \tfrac{1}{v_r^2} + \tfrac{1}{\sigma^2} & \tfrac{1}{v_1^2} & \cdots & \tfrac{1}{v_N^2} \\ 
\tfrac{1}{v_1^2} & \tfrac{1}{v_1^2} & \cdots & 0 \\ 
\vdots & \vdots & \ddots & \vdots \\ 
\tfrac{1}{v_N^2} & 0 & \cdots & \tfrac{1}{v_N^2}
\end{bmatrix} 
\begin{bmatrix}
w \\ 
z_1 \\ 
\vdots \\ 
z_N
\end{bmatrix} 
\right\rbrace \mathrm{d}\bz \mathrm{d}w \\[0.3cm]
\end{equation}
Note that the integrand from \eqref{GaussInt3} does have the full form of the multivariate Gaussian density with the specific precision matrix given above. To identify this we need to find the closed-form covariance matrix from the precision matrix and if the determinant of the covariance matrix is given by $c^2/(2\pi)^{N+1}$. Write the precision matrix as block matrix such that $A = \sum_{r = 1}^N \tfrac{1}{v_r^2} + \tfrac{1}{\sigma^2}$, $B = \Big[\tfrac{1}{v_1^2} \cdots \tfrac{1}{v_N^2}\Big]$, $ C = B^T$ and $D = \diag\Big(\tfrac{1}{v_1^2}, \cdots, \tfrac{1}{v_N^2} \Big)$. Use the partitioned matrix inversion lemma \citep[][equation 17.44]{strang:1997}  to get the blocks, $(A - BD^{-1}C)^{-1} = \sigma^2$, $(BD^{-1}C - A)^{-1}BD^{-1} = -\sigma^2 [1 \cdots 1]$, $D^{-1}C(BD^{-1}C - A)^{-1} = -\sigma^2 [1 \cdots 1]^T$ and $D^{-1} + D^{-1}C(A - BD^{-1}C)^{-1}BD^{-1}$ where its main diagonal equals to $[v_1^2 + \sigma^2, \cdots, v_N^2 + \sigma^2]$ and all off-diagonal elements are given by $\sigma^2$. Put everything together to have the covariance matrix 
\begin{equation} \label{fullCov}
\begin{bmatrix}
\sigma^2 & -\sigma^2 & \cdots & -\sigma^2 \\ 
-\sigma^2 & v_1^2 + \sigma^2 & \cdots & \sigma^2 \\ 
-\sigma^2 & \vdots & \ddots & \vdots \\ 
-\sigma^2 & \sigma^2 & \cdots & v_N^2 + \sigma^2
\end{bmatrix} 
\end{equation}
whose determinant equals to $ 1/[\det(D)\det(A-BD^{-1}C)] = \sigma^2 \prod_{r = 1}^{N} v^2_r = c^2/(2\pi)^{N+1}$ by the partitioned matrix determinant lemma. Finally, in \eqref{GaussInt3}, interchange the order of integration with Fubini-Tonelli theorem \citep{folland:2013} and integrate w.r.t. $w$ to get
\begin{equation} \label{GaussInt4}
\int\limits_{-\infty}^{\mu - m_N} \cdots \int\limits_{-\infty}^{\mu - m_1}  \mathcal{N} \left(\begin{bmatrix}
z_1 \\ 
\vdots \\ 
z_N
\end{bmatrix} \Bigg| 
\begin{bmatrix}
0 \\
\vdots \\
0
\end{bmatrix}, 
\begin{bmatrix}
v_1^2 + \sigma^2 & \cdots & \sigma^2 \\ 
\vdots & \ddots & \vdots \\ 
\sigma^2 & \cdots & v_N^2 + \sigma^2
\end{bmatrix} \right) \mathrm{d}z_1 \cdots \mathrm{d}z_N
\end{equation}
that equals to
\begin{equation} \label{GaussInt5}
F_N \left(\begin{bmatrix}
\mu \\ 
\vdots \\ 
\mu
\end{bmatrix} \Bigg| 
\begin{bmatrix}
m_1 \\
\vdots \\
m_N
\end{bmatrix}, 
\begin{bmatrix}
v_1^2 + \sigma^2 & \cdots & \sigma^2 \\ 
\vdots & \ddots & \vdots \\ 
\sigma^2 & \cdots & v_N^2 + \sigma^2
\end{bmatrix} \right)
\end{equation}
and therefore the equality \eqref{GaussInt} holds. For $N = 1$ the result follows the same as in \cite{Rasmussen+Williams:2006}.
\end{proof}

\begin{lem} \label{lemma:3}
Let $\Phi(\cdot)$ be the standard-Gaussian cumulative distribution function and denote by $\mathcal{N} (\cdot|\bmu_N, \Sigma)$ the $N$-dimensional Gaussian density function with mean parameter $\bmu_N$ and covariance matrix $\Sigma$. Then the following holds true,
\begin{equation} \label{AnGaussInt}
\int\limits_{\mathbb{R}^N}\prod\limits_{r = 1}^{N}\Phi\left(\dfrac{x_r-m_r}{v_r}\right)\mathcal{N}(\x|\bmu_N, \Sigma) \mathrm{d} \hspace{-0.04cm}\x = F_N\left(\bmu_N \vert \mathbf{m}_N, \diag(\vb_N)^2 + \Sigma \right) 
\end{equation}
where $F_N(\cdot|\mathbf{c}, \mathcal{C})$ is the $N$-dimensional Gaussian cumulative distribution function with parameters $(\mathbf{c}, \mathcal{C})$. Furthermore, $\mathbf{m}_N = [m_1  \cdots  m_N]^T \in \mathbb{R}^N$, $\bmu_N = [\mu_1 \cdots \mu_N]^T \in \mathbb{R}^N$, $\vb_N = [v_1 \cdots v_N]^T \in \mathbb{R}^N_+$ and $\Sigma$ is a covariance matrix.
\end{lem}
\begin{proof}
Let's rewrite the left-hand side of \eqref{AnGaussInt} in full and use the following transformation $[x_1, \cdots, x_N, y_1, \cdots, y_N]^T =  [w_1 + \mu_1, \cdots, w_N + \mu_N, z_1 + w_1 + m_1, \cdots, z_N + w_N + m_N]^T$. From this we note that the Jacobian of the transformation simplifies to $|\partial$ $(x_1, \cdots, x_N, y_1, \cdots, y_N)$ $/\partial (w_1, \cdots, w_N, z_1, \cdots, z_N)| = 1$. Therefore we find that 
\begin{equation} \label{AnGaussInt1}
\int\limits_{\mathbb{R}^N} \int\limits_{-\infty}^{\mu_N - m_N} \cdots \int\limits_{-\infty}^{\mu_1 - m_1}\mathcal{N}(\mathbf{w}|-\mathbf{z}, \diag(\vb_N)^2)\mathcal{N}(\mathbf{w}|\0, \Sigma)\mathrm{d}\mathbf{z} \mathrm{d}\mathbf{w}
\end{equation}
where $\mathbf{z} = [z_1 \cdots z_N]^T$ and $\mathbf{w} = [w_1 \cdots w_N]^T$. Note that the product of two multivariate Gaussians is another unnormalized multivariate Gaussian \citep[see][for example]{Rasmussen+Williams:2006}. Therefore we write 
\begin{equation} \label{AnGaussInt2}
\int\limits_{\mathbb{R}^N} \int\limits_{-\infty}^{\mu_N - m_N} \cdots \int\limits_{-\infty}^{\mu_1 - m_1}\mathcal{N}(\mathbf{z}|\0, \diag(\vb_N)^2 + \Sigma) \mathcal{N}(\mathbf{w}|c, C)\mathrm{d} \mathbf{z} \mathrm{d} \mathbf{w}
\end{equation}
where $c = - C [\diag(\vb_N)^2]^{-1} \mathbf{z}  $ and $C = ([\diag(\vb_N)^2]^{-1} + \Sigma^{-1})^{-1} $.
Interchange the order of integration with Fubini-Tonelli theorem \citep{folland:2013} and integrate w.r.t $\mathbf{w}$ to get that, 
\begin{equation} \label{AnGaussInt3}
\int\limits_{-\infty}^{\mu_N - m_N}\cdots\int\limits_{-\infty}^{\mu_1 - m_1} \mathcal{N}(\mathbf{z}|\0, \diag(\vb_N)^2 + \Sigma)\mathrm{d}\mathbf{z} = F_N(\bmu_N \vert \mathbf{m}_N, \diag(\vb_N)^2 + \Sigma)
\end{equation}
which completes the proof.
\end{proof}

Note that for $N = 1$ the result follows the same as in \cite{Rasmussen+Williams:2006}.

\section{Gaussian integral II} \label{app:AnGaussInt}

\begin{theo}
Let $\bbf = [f_1 \cdots f_N]^T \sim \mathcal{N}(\bmu, \Sigma)$, where $\mathcal{N}(\cdot)$ is the N-dimensional Gaussian density function with mean parameter $\bmu$ and covariance matrix $\Sigma$. Suppose that, conditional on $\bbf$, we perform $N$ independent Bernoulli trials with probability $\Phi(f_r)$, $r = 1, \cdots, N$, where $\Phi(\cdot)$ is the standard-Gaussian distribution function, i.e., $Y_r|f_r$ $\stackrel{\mathrm{ind}}{\sim}$ $\mathrm{Bernoulli}$ $(\Phi(f_r))$. Instead of record the values $0$ or $1$ we use the values $-1$ and $1$, so that, each Bernoulli random variable $Y_r|f_r$ has probability mass function 
\begin{equation} \label{eq:indbern}
\pi_{_{Y_r|f_r}}(y_r|f_r) = \Phi(y_r f_r)I_{\lbrace -1, 1 \rbrace}(y_r)
\end{equation}
where $I_{A}(\cdot)$ is the indicator function of a set $A$. Hence the marginal distribution of $Y = [Y_1 \cdots Y_N]^T$ is given by
\begin{equation} \label{eq:margbern}
\pi_{_{Y}}(y_1, \cdots, y_N) = F_N(\0| - I_y \bmu, I_y \Sigma I_y + I_N)
\end{equation}
where $I_y = \diag(y_1, \cdots, y_N)$, $I_N$ is the $N \times N$ identity matrix and $F_N(\cdot|\mathbf{c}, \mathcal{C})$ is the $N$-dimensional Gaussian cumulative distribution function with mean parameter $\mathbf{c}$ and covariance matrix $\mathcal{C}$.
\end{theo}

\begin{proof}
First consider the transformation $\mathbf{z}$ $=$ $I_y \f$ with $\mathbf{z}$ $=$ $[z_1 \cdots z_N]^T$ and Jacobian $|\partial(f_1, \cdots, f_N)/\partial(z_1, \cdots, z_N)|$ $=$ $1/\prod_{r = 1}^N y_r$ where we note that the absolute value of the Jacobian is 1 for any $\y \in \lbrace -1, 1 \rbrace^N$. By the change of variables method, the marginal distribution can be written as follows
\begin{align} \label{eq:intmargbern}
\pi_Y(y_1, \cdots, y_N) &= \int\limits_{\mathbb{R}^N} \pi_{Y|\bbf}(\mathbf{z}) \pi_{\bbf}(I_y^{-1}\mathbf{z}) \mathrm{d} \mathbf{z} \nonumber \\
 &= \int\limits_{\mathbb{R}^N} \prod\limits_{r = 1}^{N}\Phi(z_r) \mathcal{N}(I_y^{-1}\mathbf{z}|\bmu, \Sigma) \mathrm{d} \mathbf{z} \nonumber \\
&= \int\limits_{\mathbb{R}^N} \prod\limits_{r = 1}^{N}\Phi(z_r) \dfrac{\exp \left(-\tfrac{1}{2}(\mathbf{z} - I_y\bmu)^T (I_y \Sigma I_y)^{-1}(\mathbf{z} -I_y\bmu) \right)}{(2\pi)^{N/2} [\det (I_y \Sigma I_y)]^{1/2}} \mathrm{d} \mathbf{z} \nonumber \\
&= \int\limits_{\mathbb{R}^N} \prod\limits_{r = 1}^{N}\Phi(z_r) \mathcal{N}(\mathbf{z}|I_y\bmu, I_y\Sigma I_y) \mathrm{d} \mathbf{z}
\end{align}
where have used that $\det (I_y \Sigma I_y) = \det \Sigma$. Now, using Lemma \ref{lemma:3} yields
\begin{align} \label{eq:intmargbern1}
\pi_Y(y_1, \cdots, y_N) &= F_N(I_y\bmu|\0, I_N + I_y\Sigma I_y) \nonumber \\
&= F_N(\0|- I_y \bmu, I_N + I_y\Sigma I_y)
\end{align}
which completes the proof. 
\end{proof}

\begin{figure}[htb!]
\setlength{\parindent}{1.8cm}
\includegraphics[height = 5.3cm, width = 12.5cm]{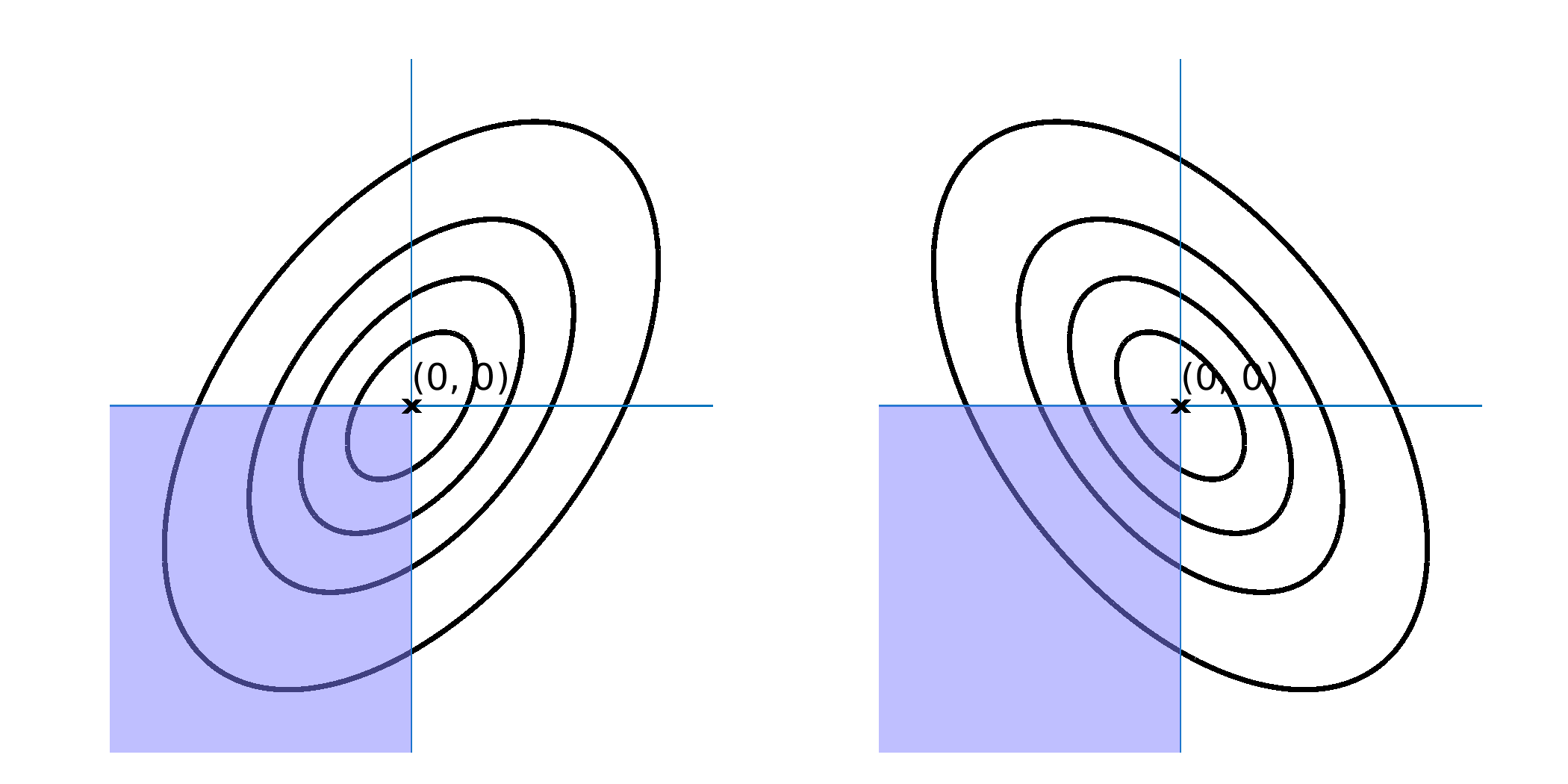}
\caption{The shaded region ilustrates the fixed region of integration. On the left-hand side the integrand is the $2$-dimensional Gaussian density function with null mean, unity variances and correlation $1/2$ and the integral corresponds to the probability $\pi(1, 1)$ or $\pi(-1, -1)$. On the right side, the region of integration is the same but the integrand is now the $2$-dimensional Gaussian with null mean, unity variances and correlation $-1/2$. This integral corresponds to the probability $\pi(-1, 1)$ or $\pi(1, -1)$.}
\label{fig:fig_5}
\end{figure}

As an example, suppose $N = 2$. Take $\bmu = [0 \ 0]^T$, $\sigma_1^2 = \sigma_2^2 \rightarrow 0$ and $\sigma_{12} = \tfrac{1}{2}$ (correlation). Therefore we have 
\begin{equation} \label{eq:intmargbern2}
\pi(y_1, y_2) = F_2\left(
\begin{bmatrix}
0 \\ 
0
\end{bmatrix} \Bigg|
\begin{bmatrix}
0 \\ 
0
\end{bmatrix},
\begin{bmatrix}
1 & \tfrac{1}{2} y_1 y_2 \\ 
\tfrac{1}{2} y_1 y_2  & 1
\end{bmatrix} \right)
\end{equation}
where, from \cite{miwa:2003}, we known that $\pi(1, 1) = \pi(-1, -1) = 1/3$ and $\pi(1, -1) = \pi(-1, 1) = 1/6$. Figure \ref{fig:fig_5} ilustrates the fixed region of integration of a $2$-dimensional Gaussian density. The integration of the $2$-dimensional Gaussian density over the shaded region correspond to the above mentioned probabilities.

\newpage
 
\bibliographystyle{apalike}
\bibliography{refs}

\appendix

\renewcommand{\thesection}{A\arabic{section}}
\renewcommand{\thefigure}{A\arabic{figure}}
\renewcommand{\thetable}{A\arabic{table}}
\renewcommand{\theequation}{A\arabic{equation}}
\setcounter{section}{0}
\setcounter{figure}{0}
\setcounter{table}{0}
\setcounter{equation}{0}

\end{document}